\documentclass[prd,twocolumn,aps,showpacs]{revtex4}

\newcommand{\be}{\begin{equation}}
\newcommand{\ee}{\end{equation}}
\newcommand{\bn}{\begin{eqnarray}}
\newcommand{\en}{\end{eqnarray}}

\def\l{\label}

\begin{document}

\title{Electromagnetic waves, gravitational coupling and duality analysis}
\author{E. M. C. Abreu$^{a}$, S. A. Diniz$^{b}$, C. Pinheiro$^{a}$ and F. C. Khanna$^{c}$}
\affiliation{$\mbox{}^{a}$Departamento de F\'{\i}sica, Universidade Federal Rural do Rio de Janeiro\\
BR 465-07, 23851-180, Serop\'edica, Rio de Janeiro, Brazil\\
{\sf E-mail: evertonabreu@ufrrj.br and carlospinheiro@ufrrj.br}\\
$\mbox{}^{b}$Centro de Ci\^encias Exatas, Universidade Federal do Esp\'{\i}rito Santo, \\
Av. Fernando Ferrari s/n$^{\underline{0}}$,Campus da Goiabeiras 29060-900 Vit\'oria ES -- Brazil\\
{\sf E-mail: diniz@cce.ufes.br}\\
$\mbox{}^{c}$Dept. of Physics, University of Alberta,\\
Edmonton, AB T6G2J1, Canada and TRIUMF, 4004, Wesbrook Mall,
V6T2A3, Vancouver, BC, Canada\\
{\sf E-mail: khanna@phys.ualberta.ca}}
\date{\today}

\begin{abstract}
In this letter we introduce a particular solution for parallel electric and magnetic fields, in a gravitational background, which satisfy free-wave equations and the phenomenology suggested by astrophysical plasma physics.
These free-wave equations are computed such that the electric field does not induce the magnetic field and vice-versa.
In a gravitational field, we analyze the Maxwell equations and the corresponding electromagnetic waves.  A continuity equation is presented.  A commutative and noncommutative analysis of the electromagnetic duality is described.  
\end{abstract}

\maketitle

\section{Introduction}
\setcounter{footnote}{0}
Plasma and Astrophysical Plasma physicists support the possible existence of
electromagnetic stationary waves with parallel $\vec{E}$ and
$\vec{B}$ fields and consequently having a null Poynting
vector \cite{um,dois,tres,quatro,cinco,seis,sete,oito}. K.R. Brownstein \cite{um} points out that
these waves may emerge as solutions of the vector equation 
$\nabla \times \vec{V}\, = \,k\, \vec{V}$ ($\vec{V}$ is a vector field
and $k$ is a positive constant). Brownstein considered
this equation for the vector potential $\vec{A}$ as
\begin{equation} \l{1.1}
{\nabla}\times \vec{A}=k\vec{A}\ ,
\end{equation}
with a particular solution
\begin{equation} \l{1.2}
\vec{A}=a\left[\vec{i}\sin kz +\vec{j}\cos kz \right]\cos \omega t
\end{equation}
and takes the associated electric and magnetic field as
\begin{equation} \l{1.3}
\vec{E}=-\ \frac{1}{c}\ \frac{\partial \vec{A}}{\partial t}=ka
\left[\vec{i}\sin kz+\vec{j}\cos kz\right]\sin \omega t \,
\end{equation}
and
\begin{equation} \l{1.4}
\vec{B}={\nabla}\times \vec{A}=ka\left[\vec{i}\sin kz\,+\,\vec{j}\,\cos kz\right]\cos \omega t\,\,.
\end{equation}
The Brownstein fields (\ref{1.3}) and (\ref{1.4}) satisfy Maxwell equations and the
usual vacuum free-wave equation. Moreover, fields, $\vec{E}$ and
$\vec{B}$ and the vector potential, $\vec{A}$, are parallel
everywhere. From (\ref{1.3}) and (\ref{1.4}) it is easy to notice the assertives above, that is, that the associated electromagnetic wave has a null Poynting
vector and, so, does not propagate energy and that the behavior of this wave
is just like the phenomenology suggested by Plasma and Astrophysical 
Plasma Physics.

In this letter, we consider the electromagnetic and gravitational
coupling and we analyze, in a gravitational background, the
corresponding Maxwell equations. With these results, we can see that with the gravitational coupling, the electromagnetic duality is broken in a commutative as well as in a noncommutative regime.
We obtain the associated free-wave equations for fields $\vec{E}$ and $\vec{B}$ and discuss the
corresponding static case. 
We will see that in a gravitational background the electric field $\vec{E}$ does not induce
the magnetic field $\vec{B}$ and vice-versa.
And we show that, in a particular situation, the electric field $\vec{E}$
may depend on time but does not induce a magnetic field $\vec{B}$, with
or without sources. Finally we show particular electrostatic
and magnetostatic solutions for Maxwell equations without sources in a
gravitational background, such  that the electric
field, $\vec{E}_0$ and the magnetic field, $\vec{B}_0$, both satisfy
the corresponding free-wave equations. Moreover, these fields 
behave  as the phenomenology in Plasma Astrophysics has suggested, that is,
$\vec{E}_0$ and $\vec{B}_0$ are parallel fields. Thus, the associated
electromagnetic wave is stationary, has a null Poynting vector and
does not propagate energy.
\\  
$\mbox{}\,\,\,\,\,\,$The work is organized as: in section $2$ we analyze the gravitational coupling in Maxwell's 
equations and the electromagnetic duality breaking.  In section 3 we formulate the solutions 
of the Maxwell equations with and without sources embedded in this gravitational background.  The conclusions are depicted in section 4.

\section{The Gravitational Coupling and the electromagnetic duality}\setcounter{equation}{0}

\paragraph*{}
The action for the gravitational and electromagnetic coupling is
written as 
\begin{equation} \l{2.1}
{\cal S}=\int d^4 x\,\sqrt{-g}\left(-\ \frac{1}{4}\ F_{\mu \nu}F^{\mu \nu}\right)
\ \,\,,
\end{equation}
requiring stationary action $(\delta S=0)$ the corresponding
Maxwell inhomogeneous equations in a gravitational background are
\begin{equation} \l{2.2}
{\cal D}_{\mu}F^{\mu \nu}={\cal J}^{\nu}\ ,
\end{equation}
where the covariant derivative above is given as:
\begin{equation} \l{2.3}
{\cal D}_{\mu}F^{\mu \nu}=\partial_{\mu}F^{\mu \nu}+
\Gamma^{\mu}_{\mu \lambda}\,F^{\lambda \nu}+\Gamma^{\nu}_{\mu \lambda}\,F^{\mu \lambda}=
{\cal J}^{\nu} \ \,\,.
\end{equation}
The second term $\Gamma^{\mu}_{\mu \lambda}$ of the covariant
derivative equation (2.3) can be written as
\begin{equation} \l{2.4}
\Gamma^{\mu}_{\mu \lambda}=-\ \frac{\partial h}{\partial x^{\lambda}}
\end{equation}
where
\begin{equation} \l{2.5}
h=\ln \sqrt{-\tilde{g}}
\end{equation}
and $\tilde{g}$ is the metric determinant.

The corresponding homogeneous Maxwell equations are
\begin{equation} \l{2.6}
{\cal D}_{\mu}F_{\nu \rho}+{\cal D}_{\nu}F_{\rho \mu}+
{\cal D}_{\rho}F_{\mu \nu}=0\ .
\end{equation}
The connection terms of this equation cancel each other in such a
way that this equation is the usual homogeneous Maxwell equation
\begin{equation} \l{2.7}
\partial_{\mu}F_{\nu \rho}+\partial_{\nu}F_{\rho \mu}+\partial_{\rho}
F_{\mu \nu}=0\ .
\end{equation}
Finally, to analyze these equations we adopt the F.R.W. cosmological metric
\begin{eqnarray} \l{2.8}
dS^2 &=&dt^2-\left(a(t)\right)^2\left\{\left(1-Ar^2\right)^{-1}dr^2 \right. \nonumber \\
& &\left. +\,r^2d\theta^2+ r^2\sin^2\theta d\varphi^2\right\}\ ,
\end{eqnarray}
where the term $a(t)$ is the scale factor and the constant $A$ may
assume values $A=1,0,-1$. Each value represents the associated
curvature of F.R.W. spatial metric.

In terms of the electric field, $E^i=F^{0i}$, and the magnetic field
$B^i=\varepsilon^{ijk}F_{jk}$, the Maxwell eq. (\ref{2.3}) and (\ref{2.6})
are explicitly given by
\begin{equation} \l{2.9}
{\nabla}\cdot \vec{E}\,=\,\rho\,{(\vec{x})}+{\nabla}h\cdot
\vec{E}\ ,
\end{equation}
\begin{equation} \l{2.10}
{\nabla}\cdot \vec{B}=0\ ,
\end{equation}
\begin{equation} \l{2.11}
{\nabla}\times \vec{E}=-\ 
\frac{\partial \vec{B}}{\partial t}\ ,
\end{equation}
\begin{equation} \l{2.12}
{\nabla}\times \vec{B}=\vec{J}(\vec{x})+\frac{\partial \vec{E}}
{\partial t}-\frac{\partial h}{\partial t}\
\vec{E}+{\nabla}h\times \vec{B}\ .
\end{equation}
Taking the divergence of equation (2.12) and using eq. (2.9) we get
the continuity equation in a gravitational background
\begin{equation} \l{2.13}
\frac{\partial \rho}{\partial t}-\frac{\partial h}{\partial t} \ \rho
+ {\nabla}\cdot \vec{J}-{\nabla}h\cdot \vec{J}=0 \;\; ,
\end{equation}
which is expressed in a covariant way as
\begin{equation} \l{2.14}
{\cal D}_{\mu}\mathcal{J}^{\mu}=0 \ ,
\end{equation}
where the covariant derivative is ${\cal
D}_{\mu}=\partial_{\mu}-\partial_{\mu}h$. 

Without electromagnetic sources $(\rho =0\ ; \ \vec{J}\,=\,0)$ the free
electromagnetic field equations in a gravitational background are:
\begin{equation} \l{2.15}
{\nabla}\cdot \vec{E}={\nabla}h\cdot \vec{E}\ ,
\end{equation}
\begin{equation} \l{2.16}
{\nabla} \cdot\vec{B}=0 \ ,
\end{equation}
\begin{equation} \l{2.17}
{\nabla}\times \vec{E}=-\frac{\partial \vec{B}}{\partial t}\ ,
\end{equation}
\begin{equation} \l{2.18}
{\nabla}\times \vec{B}=\frac{\partial \vec{E}}{\partial t}-
\frac{\partial h}{\partial t}\ \vec{E}+{\nabla}h\times \vec{B}\ \,\,.
\end{equation}
We can see clearly in eqs. (\ref{2.15})-(\ref{2.18}), that the duality invariance 
\bn
&& \vec{E} \:\rightarrow\: \vec{B} \nonumber \\
&& \vec{B} \:\rightarrow\: -\,\vec{E} \,\,,
\en
is no longer true, i.e., the gravitational coupling breaks the usual electromagnetic duality invariance of Maxwell's equations.   However, as stressed by G. F. R. Ellis and H. van Elst \cite{ee}, this is not an effect of the gravitational coupling (space-time curvature) itself, but rather due to a coordinate dependent definition of the electric and magnetic fields.  A covariant definition of the electric and magnetic fields can be performed and the result is a duality preserving form of Maxwell's source free equations \cite{ee}.

In the noncommutative case analyzed by Y. Abe {\it et al} \cite{abt} it can be demonstrated that the electromagnetic duality analyzed there is also broken under a gravitational coupling described in (\ref{2.1}) using the usual Moyal product.  This demonstration is beyond the scope of this letter. 

Taking the curl of equation (\ref{2.17}), using equations (\ref{2.15}) and
(\ref{2.18}) and, since ${\nabla}h$ does not depend explicitly on time,
we get the free-wave equation in a gravitational background for the
electric field:
\begin{eqnarray} \l{2.19}
& &\nabla^2\vec{E}-\frac{\partial^2\vec{E}}{\partial t^2} \nonumber \\
&=&{\nabla} ({\nabla}h\cdot \vec{E})-
\frac{\partial \left(\displaystyle{\frac{\partial h}{\partial t}}\vec{E}\right)}
{\partial t}-{\nabla}h\times ({\nabla}\times \vec{E})\,\,. \nonumber \\
& & \mbox{}
\end{eqnarray}

Similarly, taking the curl of eq. (\ref{2.18}), and using eq. (\ref{2.16}) and, since 
$\displaystyle{\frac{\partial h}{\partial t}}$ does not depend
explicitly on the spatial coordinates, we get the free-wave equation
in a gravitational background for the magnetic field:
\begin{equation} \l{2.20}
\nabla^2\vec{B}-\frac{\partial^2\vec{B}}{\partial t^2}=-{\nabla}
\times ({\nabla}h\times \vec{B})-\frac{\partial h}{\partial t}\ 
\frac{\partial\vec{B}}{\partial t}\ \,\,.
\end{equation}
From both equations above (\ref{2.19}) and (\ref{2.20}) we conclude that it is possible to formulate a situation in a gravitational background in which the electric field $\vec{E}$ does not induce the magnetic field and vice-versa.

\section{Solutions of the Maxwell Equations in a Gravitational
Background}\setcounter{equation}{0} 

\paragraph*{}
Now we  analyze the necessary conditions to obtain
electrostatic and magnetostatic solutions in the gravitational
background, that is, the conditions that the electric field $\vec{E}$
and the magnetic field $\vec{B}$ must satisfy in order that the
electric field $\vec{E}$ does not induce the magnetic field $\vec{B}$
and vice-versa.  Firstly, we analyze the electromagnetic fields with sources
and next in a source free system.

\subsection{With sources}
\paragraph*{}

From eq. (\ref{2.11}) it is clear that the gravitational field does not modify the
Faraday law and so it is simple to conclude that the magnetic field
$\vec{B}$ must be stationary in order not to induce an electric field
$\vec{E}$. 

On the other hand, as the Amp\`ere Law (\ref{2.12}) has been
modified by gravitation, it is interesting to see that even if the electric
field is stationary $\left(\displaystyle{\frac{\partial
\vec{E}}{\partial t}}=0\right)$ it may induce a magnetic field
$\vec{B}$ from the gravitation term 
$-\displaystyle{\frac{\partial h}{\partial t}}\ \vec{E}$. For
example, if we have a non-null stationary electric field 
\be \l{3.1}
\frac{\partial \vec{E}}{\partial t}=0\ \:\:; \qquad \vec{E}\neq 0 
\ee
and no current density,i.e., $\vec{j}=0$, eq. (\ref{2.18})
becomes 
\be \l{3.2}
{\nabla}\times\, \vec{B}\,=\,-\,\ \frac{\partial\,h}{\partial t}\,\vec{E}\,+\,{\nabla}h\,\times\,\vec{B}
\ee
and so it implies that we necessarily have a non-null magnetic field $\vec{B}$,
since $\vec{B}=0$ is not solution for this equation. 

From eq. (\ref{2.12}) it is clear that even if the electric field $\vec{E}$
depends on time it shall not induce a magnetic field $\vec{B}$ provided the condition below is satisfied:
\begin{equation} \l{3.3}
\frac{\partial \vec{E}}{\partial t}-\ \frac{\partial h}{\partial t}\ 
\vec{E}=0\ .
\end{equation}
From condition (\ref{3.3}) and from eq. (\ref{2.9}) it is simple to verify that
the electric field $\vec{E}(\vec{r},t)$ and the charge density $\rho
(\vec{r},t)$ can be written as:
\begin{equation} \l{3.4}
\vec{E}(\vec{r},t)\,=\,(a(t))^3\,\vec{E}_0(\vec{r})\ ,
\end{equation}
and
\begin{equation} \l{3.5}
\rho \,(\vec{r},t)\,=\,(a(t))^3\,\rho_0(\vec{r})\ ;
\end{equation}
where $\vec{E}_0(\vec{r})$ is a stationary vector field and
$\rho_0(\vec{r})$ is stationary ``charge density''. Furthermore the 
current density $\vec{J}$ must satisfy the equation:
\begin{equation} \l{3.6}
{\nabla}\cdot \vec{J}-{\nabla}h\cdot \vec{J}=0\;\;.
\end{equation}
If the conditions (\ref{3.4})-(\ref{3.6}) are verified and if the magnetic field
$\vec{B}$ is static, the electrostatic and magnetostatic field-equations become
\begin{equation} \l{3.7}
{\nabla}\cdot \vec{B}=0 \ ,
\end{equation}
\begin{equation} \l{3.8}
{\nabla}\times \vec{B}=\vec{J}\,+\,{\nabla}h\,\times \vec{B}\ ,
\end{equation}
\begin{equation} \l{3.9}
{\nabla}\cdot \vec{E}_0=\rho_0(\vec{r})\,+\,{\nabla}h\cdot 
\vec{E}_0\ ,
\end{equation}
\begin{equation} \l{3.10}
{\nabla}\times \vec{E}_0=0 \;\; .
\end{equation}

Any solution of equation (\ref{3.6})-(\ref{3.8}) for the charge current density
$\vec{J}(\vec{r},t)$ and for the magnetic field, $\vec{B}(\vec{r})$, can
be combined with any solution of equations (\ref{3.4}), (\ref{3.5}), (\ref{3.9}) and
(\ref{3.10}). So, under these conditions the magnetic field, $\vec{B}(\vec{r})$,
does not induce the electric field, $\vec{E}(\vec{r},t)$, and vice-versa.
This way we call these equations magnetostatic and electrostatic, but
we must remember the electric field is not static but depends on time according to eq. (\ref{3.4}).


\subsection{Without sources}
\paragraph*{}

We now consider the Maxwell equations (\ref{2.9})-(\ref{2.12}) without sources, i.e.,
$\rho =0\ ; \ \vec{J}=0$, with the condition that the electric field
$\vec{E}(\vec{r},t)$ satisfies eq. (\ref{3.3}) and that the magnetic field
$\vec{B}$ does not depend explicitly on time. The Maxwell equations
for the electrostatic and magnetostatic field-equations without
sources in a gravitational field (\ref{3.7})-(\ref{3.10}) are given by:
\begin{equation} \l{3.11}
{\nabla}\cdot \vec{B}=0 \ ,
\end{equation}
\begin{equation} \l{3.12}
{\nabla}\times \vec{B}={\nabla}h\times \vec{B}\ ,
\end{equation}
\begin{equation} \l{3.13}
{\nabla}\cdot \vec{E}_0={\nabla}h\cdot \vec{E}_0 \ ,
\end{equation}
\begin{equation} \l{3.14}
{\nabla}\times \vec{E}_0=0 \ .
\end{equation}

A particular solution in spherical coordinates is
\bn \l{3.15}
&& \vec{E}_0(r,\theta,\varphi)=\frac{\alpha}{r\ \sin \theta}\,\vec{\varphi} \nonumber \\
&& \Longrightarrow \vec{E}(r,\theta,\varphi,t)=
\alpha \ \frac{(a(t))^3}{r\sin \theta}\ \vec{\varphi}
\en
and 
\begin{equation} \l{3.16}
\vec{B}(r,\theta,\varphi)=\beta \ \frac{r}{\sqrt{1-Ar^2}}\ \vec{\varphi}\;\;,
\end{equation}
where $\alpha$ and $\beta$ are constants and $\vec{\varphi}$ is the
unit azimuthal vector: $\vec{\varphi}\,=\,-\,\vec{i}\sin \varphi\,+\,\vec{j}\cos \varphi$.

A simple substitution show that these fields satisfy all the Maxwell
equations without sources (\ref{2.15})-(\ref{2.18}). It is interesting to point
out that these electric and magnetic fields are parallel and satisfy
the free-wave eq. (\ref{2.19}) and eq. (\ref{2.20}). The electromagnetic wave has
parallel electric and magnetic fields and a null Poynting vector.  It
is a stationary wave and it does not propagate energy as suggested by
the Astrophysical Plasma phenomenology.

The term ${\nabla}h$ and the unitary azimuthal vector $\vec{\varphi}$ are perpendicular vectors, so that
$(\vec{\varphi},{\nabla}h,\vec{\varphi}\times {\nabla}h)$
form a complete set and any vector can be written as:
\begin{equation} \l{3.18}
\vec{V}\,=\,V_{\varphi}\,\vec{\varphi}\,+\,V_h\,{\nabla}h\,+\,V_{\varphi h}
(\vec{\varphi}\times \,{\nabla}h)\;\;.
\end{equation}

The vector components $V_{\varphi}$, $V_h$ and $V_{\varphi h}$ may
depend on time and spatial coordinates. General solutions $\vec{E}(r,t)$
and $\vec{B}(r,t)$ for Maxwell equations in a gravitational
background using the ansatz (\ref{3.18}) are being considered.

\section{conclusions}

In this letter we analyzed the Maxwell equations coupled to a gravitational field in the Brownstein electromagnetic waves point of view, where the electric and the magnetic vector fields are parallel and we have a null Poynting vector, i.e, there is no propagation of energy.  

In this context, we obtained an interesting covariant form of the continuity equation and free-wave equations in a gravitational background where the electric field does not induce a magnetic field and vice-versa.  In this scenario we can conclude that  the electromagnetic duality in both commutative and noncommutative regimes are no longer valid, due to a coordinate dependent definition of the electric and magnetic fields \cite{ee}.    

We constructed the necessary conditions to obtain the electrostatic and magnetostatic solutions in the gravitational background.  This is accomplished with the electromagnetic fields in a source and in a source-free systems.  These solutions satisfy the free-wave equations.  The final electric and magnetic fields obtained are parallel, as suggested by the Astrophysical Plasma phenomenology.   With the gravitational field and the unitary azimuthal vector we formulated a complete set such that any vector can be constructed as a linear combination of the vectors of this set.


As a perspective, we can analyze that in an Astrophysical Plasma, gravitation must be taken into account in a way
that the gravitational background can break the parallelism between the fields $\vec{E}$ and $\vec{B}$. Consequently, the corresponding electromagnetic wave does not have a null Poynting vector. It is not a stationary wave and thereby propagates energy.  This is a work in progress.  Another perspective is to compute the new noncommutative Maxwell equations described in \cite{abt} in a gravitational background.  

\bigskip



E.M.C. Abreu is partially supported by Funda\c{c}\~ao de Amparo \`a Pesquisa do Estado do Rio de Janeiro (FAPERJ).
The authors acknowledge J.A. Hela\"{y}el-Neto for discussions and
comments. Thanks are also due to G.O. Pires for reading the very first manuscript.

\end{document}